# Measuring and Characterizing International Collaboration Patterns in Indian Scientific Research

**Jyoti Dua[a], Vivek Kumar Singh[a, 1], Hiran H.L[b].**

[a] Department of Computer Science, Banaras Hindu University, Varanasi-221005. (India)
[b] DST-Center for Policy Research, Indian Institute of Science, Bengaluru-560012. (India)

**Abstract:** Scientific collaboration at international level has increased manifolds during last two decades. Collaboration is not only associated with higher research productivity but has also been found to be positively correlated with impact. Considering the benefits and advantages of international collaboration for the national science and technology systems of a country, policymakers in different countries have designed programs to promote international collaboration in science. This paper attempts to measure and characterize the international collaboration patterns in Indian scientific research. Research publication data for the last 20 years (2001-2020), obtained from Web of Science, is analysed for the purpose. Analytical results show that India's international collaboration has grown at a rate of 12.27% during this period, rising from only 20.73% internationally collaborated papers in 2001 to 32.35% internationally collaborated papers in 2020. USA, Germany, England, South Korea and China are found to be the top collaborating partners for India during his period, however, some subject area-wise variations are also seen. Among the internationally collaborated papers, about 50% papers have an Indian researcher as lead (first) author, and more than 50% authors of such internationally collaborated papers are from India. The internationally collaborated papers from India are also found to have a slight advantage in terms of citation impact and social media visibility. The probable factors shaping the Indian international collaboration and the major policy implications are discussed.

## Introduction

Collaborative research has become an integral part of any flourishing academic or industrial ecosystem. Collaboration and its benefits are well-studied under different perspectives. Mutual benefits in the form of resource sharing and knowledge transfer, complementary and/or common solutions for shared problems, etc., are the major drivers of collaborative research. Several studies have analyzed collaboration at international level and have observed that it has rose linearly during the last two 2-3 decades, as measured in terms of number of internationally co-authored papers published (Glanzel, 2001; Persson, Glänzel & Danell, 2004; Lee & Bozeman 2005; Wagner & Leydesdorff, 2005; Leydesdorff & Wagner, 2008; Mattsson et al., 2008; Adams, 2012). A strong affinity between collaboration and research productivity has also been observed along with the indications that collaboration may have significant impact on citations (Glanzel, 2001; Abramo,

---

[1]Corresponding author. Email: vivekks12@gmail.com

DÁngelo, & Solazzi, 2011; Abramo, DÁngelo & Murgia, 2017). The benefits of collaboration on productivity in academic research was also given by Abramo, DÁngelo & Di Costa (2009); Ductor (2015); Parish, Boyack & Ioannidis (2018), etc. In industrial context, inventor collaborations are found to have profound effect on productivity (Favaro, Ninka & Turvani, 2012). It is to be noted that, after recognizing the benefits of international research collaborations, several national and international funding agencies have invested in policies that intended to foster collaborative research (Katz & Martin, 1997; Wagner et al., 2001; Boekholt et al., 2009; Wagner & Leydesdorff, 2005; Jeong et al., 2014). Emphasis by different countries to engineer international collaborations, indicates the vital role of collaborations in the progress of Science and Technology. Efforts to create and foster productive collaborations between nations is also a consequence of the increased awareness about the importance of collaborative research. The pursuit towards United Nation's Sustainable Development Goals[2] or SDGs calls for proactive research and development effort at global scale in different areas, which will require intense collaboration between countries.

In the knowledge-based economy, in terms of scientific, technological and other types of knowledge, every nation is a prosumer (producer as well as consumer). However, the level of their production and consumption of knowledge varies in different degrees in different fields according to their overall capacities (indigenous capacities as well as the capacities built through external and/or extramural collaboration and various other schemes). India is steadily growing as a major knowledge producer as evident from its ranking in the top order by various study reports. For instance, two studies commissioned by DST, carried out by Elsevier's Analytical Services (DST-Elsevier, 2020) and Clarivate Analytics (DST-Clarivate, 2019) placed India at 5th and 9th positions, respectively, in research output volume. According to the National Science Foundation (NSF) report on Science and Engineering indicators (2020), India is ranked at 3rd position in terms of global research output (NSF Report, 2020). Though such a leap seems to be rapid, it is a cumulative consequence of a continuum of efforts through many decades. Establishment and management of fruitful collaborative ties with reputed international institutions or agencies had been one of such vital efforts that paid well. The science and technology research in India post-independence has drawn a lot from different international collaborations. In fact, all the first-generation Indian Institutes of Technology involved collaborated approach at international level. For example, IIT Bombay (one among the institutions recognized by Govt. of India as institutions of eminence) was established in 1958 with the funding from the then USSR and assistance from UNESCO (IIT Bombay-Institute history, 2014).

Over the last two decades, there have been major shifts in global S&T landscape and India's international interactions and knowledge networks have also evolved. Considering the fact that the Reddy (2014) attributed the relatively low impact of Indian research (despite its 10th position in global share of research output) in his study over the period of 1996-2012 to the lesser share of collaborations in a scale comparable to top order countries, it is highly important to analyze whether collaborations have improved over time adequately. In a sense, the evidence or indications of improvement in collaboration can explain to some extent the leap in country's productivity. Measurement of the improvement of impact can reflect how much effective is the improvement in productivity and thereby how important role productivity has played in the progress of India's scholarly research. There lies the importance of measuring the country's international collaboration at this point of time. Even more important is the characterization or examination of the nature or characteristics of international collaborations. Specific concerns of such an examination are the identification degree/level of collaborations with countries and among different fields of research to determine top collaborating countries, top research areas/fields of collaboration, knowledge flows in such collaboration etc.

---

[2] https://sdgs.un.org/goals

All the above factors increase the relevance of studies related to collaboration that addresses the concerns such as (i) assessment of the present status of collaborative research with respect to a country/region or any other entity, and (ii) characterization of collaboration patterns with respect to that entity, etc. There lies the motivation and novelty of this research. In this work, we attempt to address these concerns for the country 'India'. Thus, the objectives of this paper are to measure and characterize international collaboration patterns in Indian scientific research during 2001-20 and understand how these collaborations have evolved recently, including discussing the relevant policy implications thereof. More precisely, the paper attempts to answer following questions:

**RQ1**: What proportion of research output of India involves international collaboration and whether the propensity to collaborate internationally has increased over time?

**RQ2**: What are the major collaborating research partner countries for India?

**RQ3:** Which subject areas witness higher international collaboration (and with which countries)?

**RQ4:** What proportion of internationally collaborated papers from India has lead/ majority of author(s) from India?

**RQ5:** Do internationally collaborated papers of India get higher impact and visibility as compared to indigenous papers?

Achievement of these objectives can be vital for the extrication of suitable policy implications to improve the overall research output and impact of India through prospective collaborations.

**Related Work**

Research collaboration involves a group of researchers working together as a team to achieve some common goals. Collaboration can involve cooperation at different levels- individuals, institutions and countries (Katz & Martin, 1997). Several studies have analyzed collaboration at international level and have observed that it has rose linearly during the last two 2-3 decades, as measured in terms of number of internationally co-authored papers published (Glanzel, 2001; Persson, Glänzel & Danell, 2004; Lee & Bozeman 2005; Wagner & Leydesdorff, 2005; Leydesdorff & Wagner, 2008; Mattsson et al., 2008; Adams, 2012). A strong affinity between collaboration and research productivity has also been observed along with the indications that collaboration may have significant impact on citations (Glanzel, 2001; Abramo, DÁngelo, & Solazzi, 2011; Abramo, DÁngelo & Murgia, 2017). The internationally collaborated papers are particularly found to attract higher citations as compared to domestically collaborated papers (Bordons et al., 1996; Glanzel & Schubert, 2001; Persson, 2010; Bote, Olmeda-Gomez & Moya-Angeon, 2013; Nguyen, Ho-Le, & Le, 2017). It has also been pointed that, countries that are highly "open", produce high impact of research; where "open" is defined as having government policies that promote international collaboration and scientific mobility (Wagner & Jonkers, 2017).

Considering the benefits and advantages of international collaboration for the national science and technology systems of a country, policymakers in different countries have designed programs to promote different kinds of international collaboration (Katz & Martin, 1997; Boekholt et al., 2009; Wagner et al., 2001). There have also been efforts to identify the main driving factors of international collaboration in scientific research. Several such major driving factors identified include: science policies (Glanzel et al., 1999), need for use of complex equipment and technical infrastructure (Beaver, 2001; Birnholtz, 2007; D'lppolito & Ruling, 2019), increasing specialization of science (Leahey & Reikowsky, 2008), need for knowledge sharing and training of scientists (Katz & Martin, 1997; Beaver, 2001) and globalization of science. The notion of proximities/ distances is another important aspect analysed in relation to international research collaboration (Boschma, 2005; Balland, Boschma & Frenken, 2015; Vieira, Cerdeira, & Teixeira, 2022).

In the broader context of the relevant factors, and benefits and advantages of international collaboration; many previous studies focused their attention on measuring and understanding international collaboration patterns of individual countries, such as for India (Basu & Kumar, 2000; Basu & Aggarwal, 2001; Gupta & Dhawan, 2003; Anuradha & Urs, 2007; Arunachalam & Viswanathan, 2008), Brazil (Leta, & Chaimovich, 2002; McManus et al., 2020), Mexico (Marmolejo-Leyva, Perez-Angon & Russell, 2015), Korea (Kim, 2005), Vietnam (Nguyen, Ho-Le, & Le, 2017), UK (Adams, Gurney, & Marshall, 2007) and Russia (Pislyakov & Shukshina, 2014) etc. Several other previous studies focused on international collaboration in different regions, such in Africa (Mêgnigbêto, 2013; Adams et al., 2014), Asian region (Arunachalam, Srinivasan, & Raman, 1994; Arunachalam & Doss, 2000; Uddin & Singh, 2014;), European region (Braun & Glaenzel, 1996; Glänzel, Schubert, & Czerwon, 1999), and BRICS countries (Singh & Hasan, 2015; Finardi, 2015; Bouabid, Paul-Hus, & Larivière, 2016; Finardi & Buratti, 2016; Leta, das Neves Machado, & Canchumani, 2019; Sokolov et al., 2019) etc.

Some previous studies have analysed the international collaboration in Indian scientific research (Basu & Kumar, 2000; Basu & Aggarwal, 2001; Gupta, Munshi & Mishra, 2002; Gupta & Dhawan, 2003; Anuradha & Urs, 2007; Arunachalam & Viswanathan, 2008). For example, Basu & Kumar (2000) analysed the extent of international collaboration in Indian science for the 1990 to 1994 period from SCI data and found an increase both in terms of output and the extent of the network. The bulk of Indian scientific co-operation was with the developed Western nations and Japan. Basu & Aggarwal (2001) analysed the data from major Indian institutions for 1997 from SCI and quantified the gain in impact through foreign collaboration. Their cluster analysis showed that different institutions had different patterns of gain from international collaboration. Gupta, Munshi & Mishra (2002) analysed the 1992-99 SCI data for India to find out how much did India collaborate with other South Asian neighbors and found varying amount of collaboration with different South Asian countries. Gupta & Dhawan (2003) analysed India's collaboration in Science & Technology with China during 1994-99 and found that the number of collaborated papers grew from 21 in 1994 to 74 in 1999. However, it was also found that S&T collaboration between India and China has been taking place mainly through multilateral channels and the output through bilateral channels was very small (11.7%). Anuradha & Urs (2007) used correspondence analysis to compute bibliometric indicators of India research collaboration patterns during 1993 to 2000.

Arunachalam & Viswanathan (2008) analysed the Indian publication data for 2000-07 period to understand the collaboration between India and China. The number of collaborated papers were found to have increased from 124 papers in 2000 to 361 in 2007, with multilateral collaboration being the main channel of collaboration.

Despite the abovementioned previous studies having analysed the international collaboration patterns in Indian scientific research, there is a need for a fresh and up to date analysis due to various reasons. *First*, most of existing studies on Indian international research collaboration are quite old, with the most recent being done about 15 years before, whereas the Science and Technology research landscape of the world has changed significantly since then. *Second*, majority of the previous studies for India are for the pre-ICT revolution era and the collaboration patterns may have changed significantly since then as ICT has diminished different kinds of distances in scientific collaboration. *Thirdly*, the existing studies lack the subject-specific collaboration pattern analysis. *Fourth*, the leadership role and knowledge flows through international collaboration of India are not known. *Fifth*, to the best of our knowledge, there is no existing analysis on impact of international collaboration on citation and social media visibility of Indian research. This article, therefore, attempts to bridge these research gaps by analyzing international collaboration patterns in Indian scientific research during 2001 to 2020. In addition to a measuring and characterizing international collaboration patterns, the article also presents a discussion on policy implications of the observations.

**Data & Method**

The Indian research output data for the period 2001-20 has been used for the bibliometric analysis. The data is download from Web of Science database core collection by using the search query: *CU=India and PY= (2001-2020)*. Publication data downloaded was restricted to document types 'Article' and 'Review', as we were interested mainly in analyzing journal publications. The data downloaded was preprocessed to remove duplicate and null values. A total of 784,157 publication records were left for further analysis. There were 60 metadata fields in the publication record data downloaded, such as author address (C1), document type (DT), publication type (PT) etc. The analysis mainly involved processing of information in DI, C1, WC and Z9 fields. The DI represent digital object identifier (DOI) and C1 denotes author's affiliation, including affiliation country. The C1 field information was used to identify publication records that involved authors only from India (domestic publications) and those that had authors from some other country as well (internationally collaborated publications). A total of 200,498 publication records were identified as internationally collaborated publications (ICP) and 583,659 publication records were domestic publications.

The publication records tagged into the two groups were then analyzed to compute different results. The year-wise ICP proportion for India was identified and plotted. The Compound Annual Growth Rate (CAGR) of the ICP instances as well as India's total output is also computed. The CAGR is computed by using the following formula:

$$\text{CAGR} = \left(\left(\frac{Vfinal}{Vbegin}\right)^{\frac{1}{t}} - 1\right) * 100$$

where, V*final* is number of publication records in the year 2020, V*begin* is the number of publication records in the year 2001, and t is the time period in years. The major collaborating countries for India and the number of collaborated papers with them was identified and plotted next. Thereafter, the subject area-wise international collaboration patterns are analysed. For this purpose, the publication records are categorized into 14 broad subject areas by using the scheme proposed in (Rupika, Uddin & Singh, 2016). These 14 broad subject areas are: Agriculture (AGR), Arts & Humanities (AH), Biology (BIO), Chemistry (CHE), Engineering (ENG), Environment (ENV), Geology (GEO), Information (INF), Material (MAR), Mathematics (MAT), Medicine (MED), Multidisciplinary (MUL), Physics (PHY), Social Science (SS). The internationally collaboration patterns in these 14 subject areas are identified and the major collaborating countries in each of these subject areas listed.

The next analysis involved identifying what proportion of India's ICP instances have an Indian author as first/ lead author. This was done by processing the author affiliation field data. Similarly, the affiliating country information for all authors in a paper was processed to find out how many of the ICP instances of India have more than 50% authors from India. The subject area differentiation for these patterns is also identified. Finally, the citation impact and social media visibility of ICP and domestic papers from India were computed and compared. The citation count was obtained from the Z9 field of the publication metadata. The social media visibility in different platforms was obtained from Altmetric.com. In order to obtain altmetric data from Altmetric.com, a DOI look up was performed for all the DOIs in the WoS data. Out of the 784,157 publication records, a total of 187,566 records are found to be covered by Altmetric.com, which is about 23.91% of the total data. The altmetric data was obtained for a total of 74,035 publications out of 200,498 ICP instances, and 113,531 publications out of 583,659 domestic publications. Altmetric.com has 46 fields in the data, including DOI, Title, Twitter mentions, Facebook mentions, News mentions, Altmetric Attention Score, OA Status, Subjects (FoR), Publication Date, URI, etc. Out of these, we mainly used data for Twitter, Facebook, and Blog platforms. The data from Altmetric.com was downloaded in the month of Aug. 2021.

**Results**

We now present the results for the analysis, organized in a manner to answer the different research questions proposed.

### *RQ1: What proportion of research output of India involves international collaboration and whether the propensity to collaborate internationally has increased over time?*

For finding an answer to this question, the proportion of domestic and ICP instances in Indian research output are identified, in a year-wise manner from 2001 to 2020. **Table 1** presents the total research output, ICP and domestic papers from India during the period 2001-2020. It can be observed that in the year 2001, a total of 2,354 papers out of total 11,357 papers from India, involved international collaboration. This constitutes 20.73% of the total papers. By the year 2020, internationally collaborated papers have increased to 23,815, which is 32.35% of the total research output of 73,614 papers by India in the year. Thus, not only the volume of internationally collaborated papers has increased, but also their proportionate share in Indian research output. For the whole 20-year period taken together, it is found that 25.57% of total research papers involve collaboration internationally.

For a clearer visualization of the trend of international collaboration of India during the period and the rate of growth of international collaboration, we plot the year-wise count of total research papers and the ICP instances together and compute their CAGRs. **Figure 1** presents the plot of year-wise total research output from India and the research output that is internationally collaborated. It can be observed that both, the total research papers from India as well as the internationally collaborated papers, have increased during the period. However, the rate of growth of ICP instances (12.27%) is higher as compared to rate of growth of total research output (9.8%). Thus, the analytical results show that during the 2001-20 period, India has about 1/4$^{th}$ of its research output involving international collaboration, and that India's propensity to collaborate internationally has increased during the period, rising from 20.73% research papers in 2001 to 32.35% research papers in 2020. This is a significant rise in proportionate share, especially considering the fact that total research output from India during this period has also increased significantly.

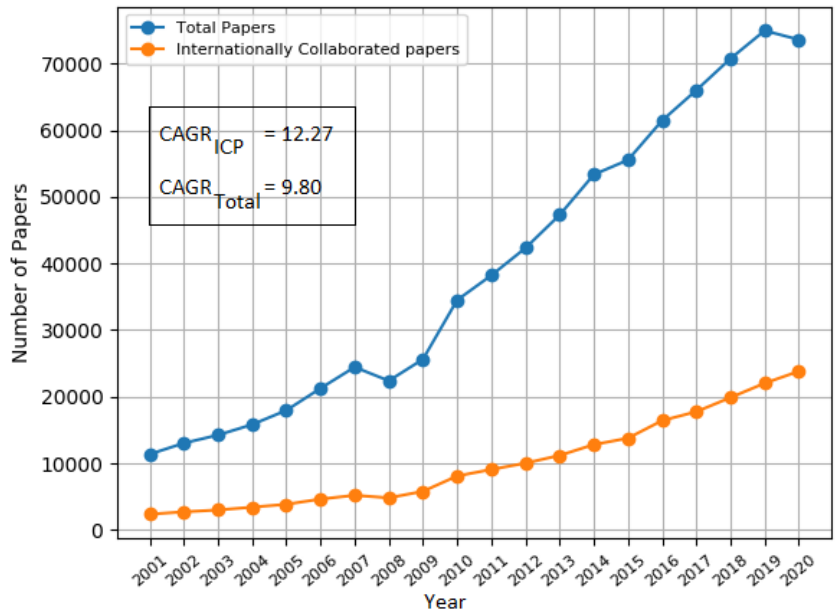

**Figure 1: Growth of India's total research output and internationally collaborated papers**

**Table 1: Number and percentage of internationally collaborated and domestic research papers for India (2001-20)**

| Year | Total No. of papers | No. and % of internationally collaborated papers | No. and % of domestic papers |
|---|---|---|---|
| 2001 | 11357 | 2354 (20.73%) | 9003 (79.27%) |
| 2002 | 13024 | 2707 (20.78%) | 10317 (79.22%) |
| 2003 | 14237 | 2986 (20.97%) | 11251 (79.03%) |
| 2004 | 15816 | 3409 (21.55%) | 12407 (78.45%) |
| 2005 | 17945 | 3838 (21.39%) | 14107 (78.61%) |
| 2006 | 21252 | 4623 (21.75%) | 16629 (78.25%) |
| 2007 | 24441 | 5188 (21.23%) | 19253 (78.77%) |
| 2008 | 22387 | 4810 (21.49%) | 17577 (78.51%) |
| 2009 | 25624 | 5784 (22.57%) | 19840 (77.43%) |
| 2010 | 34,508 | 8,089 (23.44%) | 26,419 (76.56%) |
| 2011 | 38,219 | 9,071 (23.73%) | 29,148 (76.27%) |
| 2012 | 42,328 | 10,018 (23.67%) | 32,310 (76.33%) |
| 2013 | 47,290 | 11,181 (23.64%) | 36,109 76.36%) |
| 2014 | 53,349 | 12,807 (24.01%) | 40,542 (75.99%) |
| 2015 | 55,542 | 13,771 (24.79%) | 41,771 (75.21%) |
| 2016 | 61,477 | 16,412 (26.70%) | 45,065 (73.30%) |
| 2017 | 66,016 | 17,746 (26.88%) | 48,270 (73.12%) |
| 2018 | 70,767 | 19,871 (28.08%) | 50,896 (71.92%) |
| 2019 | 74,964 | 22,018 (29.37%) | 52,946 (70.63%) |
| 2020 | 73,614 | 23,815 (32.35%) | 49,779 (67.65%) |
| **Total** | 784,157 | 200498 (25.57%) | 583659 (74.43%) |

We have also tried to measure what part of internationally collaborated papers from India involve bilateral and multilateral collaboration. Here, bilateral means all such papers where there is a collaboration between researcher(s) from India and one or more researchers from one more country. Multilateral collaboration means research papers that involve researchers from three or more countries. **Figure 2** shows a plot of year-wise proportion of bilateral and multilateral collaborated research papers from among the total internationally collaborated research papers. Out of total ICP instances, 32.79% research papers are instances of multilateral collaboration and 67.21% research papers are instances of bilateral collaboration. However, the proportion of multilateral collaborations are seen rising during the observation period.

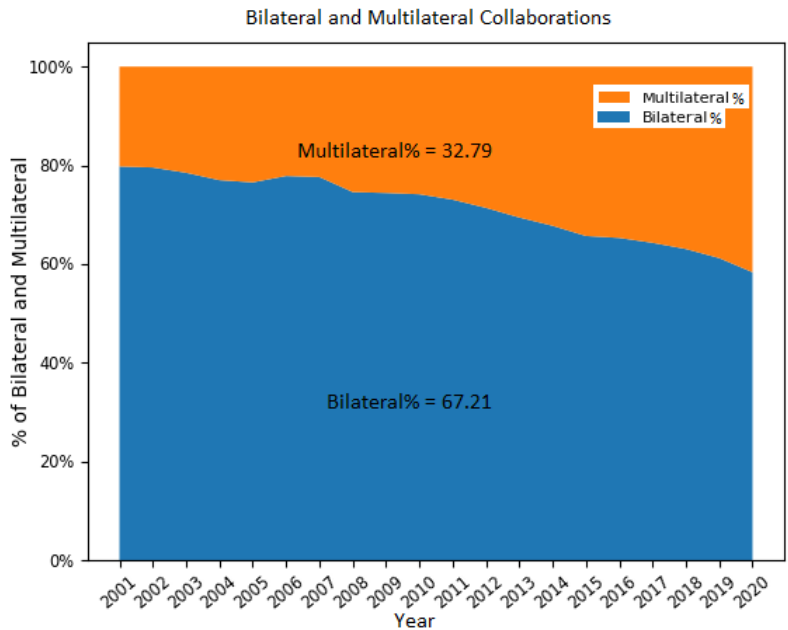

**Figure 2: Bilateral and Multilateral instances of Indian ICP**

*RQ2: What are the major collaborating research partner countries for India?*

The second research question involved identifying the major collaborating partner countries for India. For this purpose, the top countries with which India obtained highest collaborated output during the period, are identified. **Table 2** presents the number and % of ICP instances for the top 20 countries with India, grouped into two blocks of 10 year each, i.e., during 2001-10 and 2011-20. Further, the CAGR of internationally collaborated papers with each country during the period are also computed. It can be observed that, USA, Germany, England, South Korea and China are the top 5 major collaborating partners. USA accounts for the largest share of more than 30% ICP instances of India, although its share declined marginally from 34.14% in 2001-10 period to 31.22% in 2011-20 period. Germany is the 2nd largest collaborator, though here too there is a slight decline from 14.60% in 2001-2010 to 10.76% in 2011-2020. The collaborated research output with England, South Korea and China, however, has increased from 2001-10 to 2011-20 period. Among some interesting patterns to observe is Saudi Arabia that records a rapid growth in the collaborated research output with India, from 0.74% in 2001-10 period to 7.29% in 2011-20 period, with a CAGR value of 33.36%. The collaboration with South Korea, China, Australia, South Africa and Malaysia is also found to be increasing at a higher rate. Germany shows lowest CAGR value, indicating that the collaboration between India and Germany is not increasing significantly. Thus, the values in the table point to rise in India's research collaboration with newer partners, while collaboration with existing partners either remained intact or decreased a bit in terms of proportion (though volume has increased). For example, India's ICP instances with USA have increased from 14,951 papers in 2001-10 to 48,927 papers in 2011-20, an increase of more than three times. However, at the same time proportionate share of USA in India's ICP instances has decreased, indicating emergence of new collaboration partners. A similar pattern is observed for Germany. China, South Korea, Australia and Saudi Arabia are seen as new collaboration partners with increased research collaboration.

**Figure 3** shows a graphical visualization illustrating the major collaborating partner countries of India during the 2001-20 period, along with the intensity of each collaboration. The darker shade represents higher collaboration. It can be clearly observed that majority of the major collaborating partners with India are located distantly from India (USA being the farthest and yet having highest collaborated output). Only, China is a neighboring country for India, and South Korea, Malaysia and Taiwan are South East Asian neighbors of India. India's collaboration with its South Asian neighbors is found to be extremely small. Brazil and South Africa are also among the emerging international collaboration partners for India, though their proportionate share is still quite low. The 'Discussion' section of the paper presents more discussion on this aspect, including the connections with proximity theories of research collaboration.

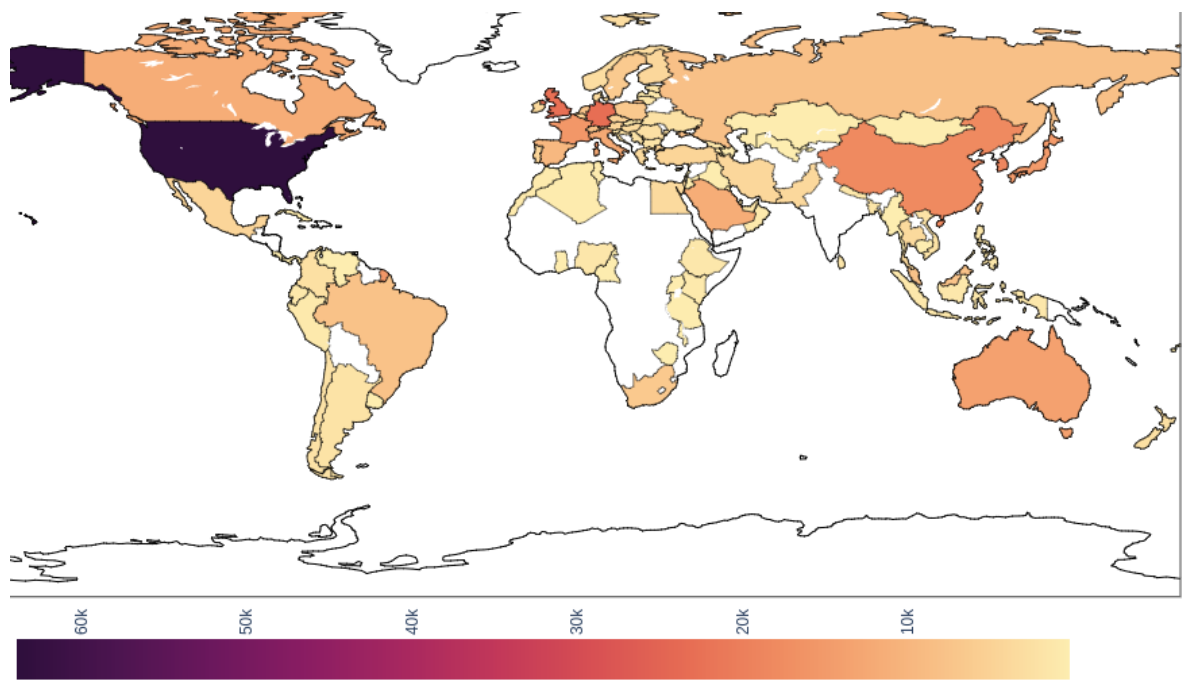

**Figure 3: Graphical visualization of major collaborating countries with India (2001-20)**

**Table 2: Top 20 Collaborating countries with India and the collaborated research output**

| Countries | Internationally Collaborated Papers | | CAGR% (2001-2020) |
|---|---|---|---|
| | 2001-10 | 2011-20 | |
| USA | 14951 (34.14%) | 48927 (31.22%) | 10.59 |
| Germany | 6395 (14.60%) | 16859 (10.76%) | 9.51 |
| England | 4520 (10.32%) | 18272 (11.66%) | 13.10 |
| South Korea | 3186 (7.28%) | 15698 (10.02%) | 19.79 |
| China | 2433 (5.56%) | 15856 (10.12%) | 18.89 |
| France | 3725 (8.51%) | 12220 (7.80%) | 10.28 |
| Japan | 4302 (9.82%) | 10807 (6.90%) | 9.81 |
| Australia | 2005 (4.58%) | 11996 (7.65%) | 17.12 |
| Italy | 2102 (4.80%) | 10393 (6.63%) | 13.09 |
| Canada | 2394 (5.47%) | 9386 (5.99%) | 12.27 |
| Saudi Arabia | 322 (0.74%) | 11425 (7.29%) | 33.36 |
| Spain | 1419 (3.24%) | 8395 (5.36%) | 14.72 |
| Taiwan | 1680 (3.84%) | 6731 (4.30%) | 14.08 |
| Russia | 1546 (3.53%) | 6722 (4.29%) | 13.65 |
| Switzerland | 1516 (3.46%) | 6718 (4.29%) | 14.30 |
| Brazil | 1142 (2.61%) | 6742 (4.30%) | 15.87 |
| Malaysia | 1179 (2.69%) | 6595 (4.21%) | 18.46 |
| Netherlands | 1443 (3.30%) | 6005 (3.83%) | 11.72 |
| South Africa | 526 (1.20%) | 6123 (3.91%) | 19.09 |
| Poland | 1106 (2.53%) | 5522 (3.52%) | 14.87 |

**Note:** Some internationally collaborated papers may involve multiple countries and hence the overall % value may be greater than 100.

*RQ3: Which subject areas witness higher international collaboration (and with which countries)?*

The subject area-wise patterns of India's international collaboration are analyzed next to understand whether the different subject areas exhibit different levels of international collaboration. Or in other words, whether research in some subject areas involve higher international collaboration as compared to others. The fourteen broad subject area grouping, as described in Methodology, is used for the purpose.

**Figure 4** shows the year-wise proportion of domestic and ICP instances for the 14 subject areas. Social Science (SS), Mathematics (MAT) and Multidisciplinary (MUL) are found to be the three subject areas with relatively higher proportion of ICP instances. The percentage of ICP for these areas are 41.85% (for SS), 35.03% (for MAT) and 31.81% (for MUL). The other subject areas with relatively better proportion of papers as ICP are PHY (29.23%), GEO (27.54%) and ENV (26.23%). Out of all the subject areas, Agriculture (AGR) has the lowest proportion of ICP, with only 20.26% of the total research papers during 2001-20 involving international collaboration. It may be noted that the $CAGR_{ICP}$ values for all subject areas are greater than $CAGR_{Non-ICP}$, suggesting that ICP proportion in each subject area is increasing. The MUL subject area, with a $CAGR_{ICP}$ of 22.38% is the one to record highest growth in internationally collaborated research output. This is followed by INF with $CAGR_{ICP}$ value of 17.41%. PHY and AGR are the two subject areas with lowest $CAGR_{ICP}$ value of 8.27%, 9.87%, respectively.

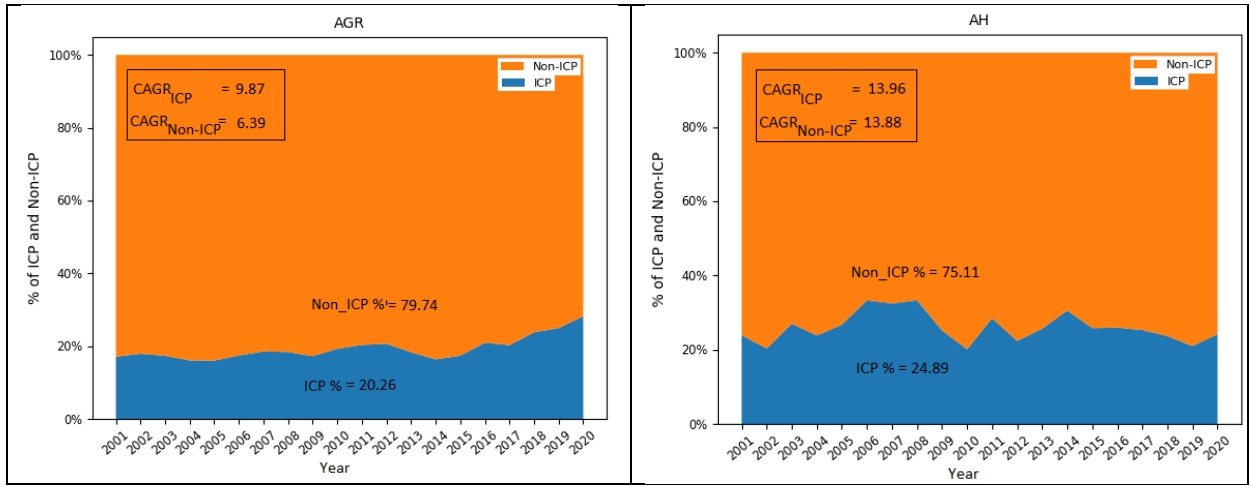

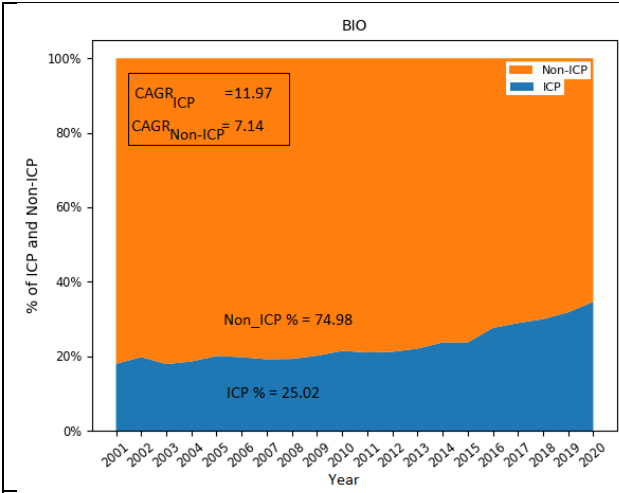
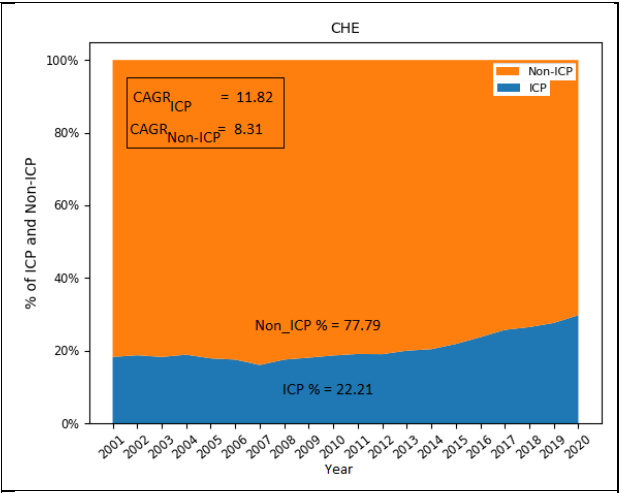
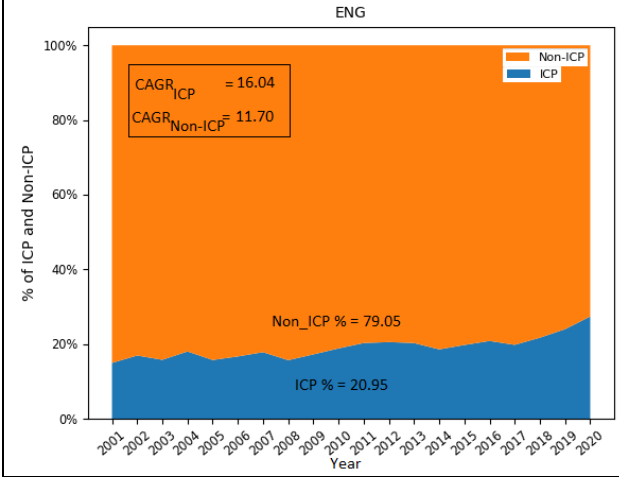
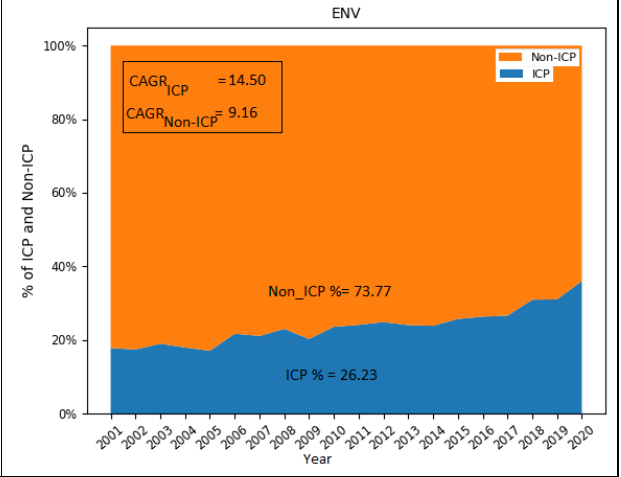
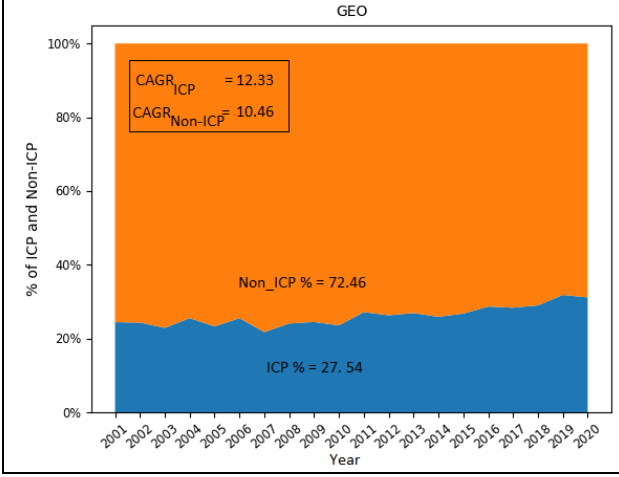
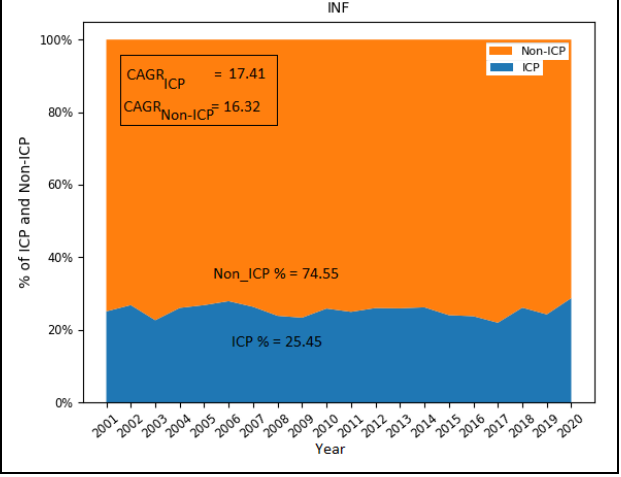

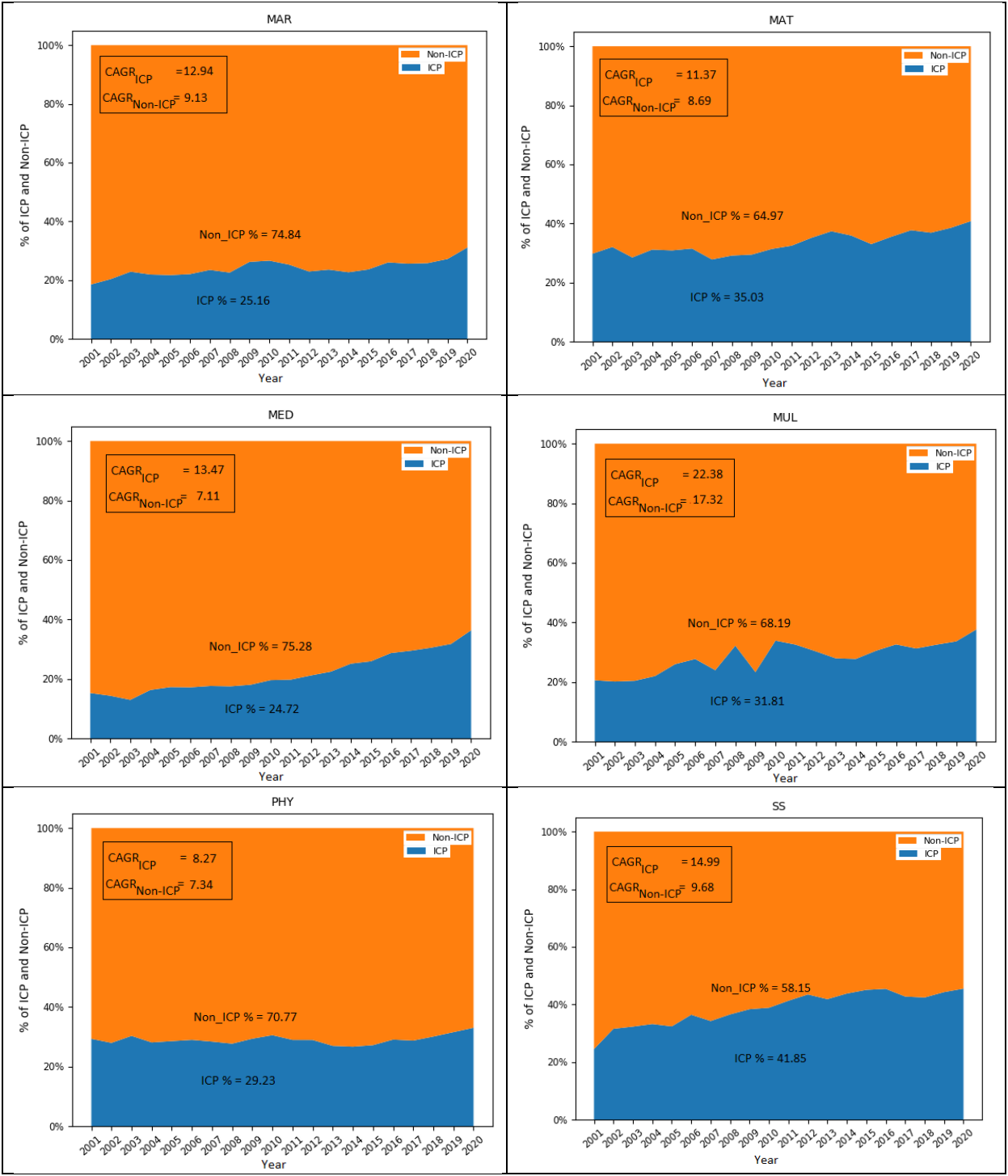

**Figure 4: ICP proportion and trends in different subject areas**

**Table 3: Top 10 collaborating countries in different subject areas**

| Subject Area | Countries and the number and percentage of internationally collaborated papers | | | | | | | | | |
|---|---|---|---|---|---|---|---|---|---|---|
| AGR | USA (2996) 25.61% | Australia (1119) 9.56% | China (978) 8.36% | Germany (922) 7.88% | South Korea (908) 7.76% | England (731) 6.25% | Japan (689) 5.89% | Saudi Arabia (561) 4.79% | Canada (530) 4.53% | France (472) 4.03% |
| AH | USA (642) 41.23% | England (185) 11.88% | China (154) 9.89% | Australia (128) 8.22% | Germany (98) 6.29% | Canada (95) 6.10% | France (87) 5.59% | Saudi Arabia (56) 3.60% | South Korea (54) 3.47% | Netherlands (52) 3.34% |
| BIO | USA (17286) 36.72% | England (4294) 9.12% | Germany (4051) 8.60% | South Korea (3652) 7.76% | China (3414) 7.25% | France (3056) 6.49% | Australia (2954) 6.27% | Japan (2795) 5.94% | Italy (2713) 5.76% | Saudi Arabia (2531) 5.38% |
| CHE | USA (10032) 21.66% | South Korea (5145) 11.11% | Germany (4280) 9.24% | Japan (3426) 7.40% | Saudi Arabia (3360) 7.26% | England (2933) 6.33% | France (2538) 5.48% | China (2517) 5.44% | Taiwan (2000) 4.32% | Spain (1999) 4.32% |
| ENG | USA (6236) 24.31% | South Korea (2698) 10.52% | China (2257) 8.80% | England (1845) 7.19% | Australia (1542) 6.01% | Saudi Arabia (1451) 5.56% | Canada (1397) 5.45% | Germany (1328) 5.18% | Malaysia (1289) 5.02% | Japan (1037) 4.04% |
| ENV | USA (4916) 27.87% | China (1821) 10.32% | Germany (1623) 9.20% | England (1588) 9.00% | Australia (1447) 8.20% | South Korea (1359) 7.70% | Japan (1286) 7.21% | France (1039) 5.89% | Canada (916) 5.19% | Saudi Arabia (901) 5.11% |
| GEO | USA (4868) 29.69% | England (1749) 10.67% | Germany (1709) 10.42% | Australia (1509) 9.20% | China (1487) 9.07% | Japan (1323) 8.07% | France (1290) 7.87% | South Korea (1126) 6.87% | Canada (1064) 6.49% | Italy (688) 4.20% |
| INF | USA (4920) 30.69% | China (2126) 13.26% | England (1242) 7.75% | South Korea (1146) 7.15% | Australia (1057) 6.59% | Canada (1050) 6.55% | Saudi Arabia (838) 5.23% | France (707) 4.41% | Germany (698) 4.35% | Malaysia (625) 3.90% |
| MAR | USA (8602) 20.42% | South Korea (5761) 13.67% | Germany (4644) 11.02% | Japan (3271) 7.76% | Saudi Arabia (2456) 5.83% | England (2405) 5.71% | China (2379) 5.65% | France (2104) 4.99% | Malaysia (1877) 4.46% | Taiwan (1652) 3.92% |
| MAT | USA (3877) 23.68% | China (1621) 9.90% | Saudi Arabia (1423) 8.69% | Germany (1078) 6.58% | South Korea (953) 5.82% | Canada (910) 5.56% | France (839) 5.12% | England (801) 4.89% | Turkey (596) 3.64% | Spain (591) 3.61% |
| MED | USA (21597) 47.93% | England (8055) 17.88% | Australia (5161) 11.45% | Canada (4450) 9.88% | Germany (3996) 8.87% | China (3951) 8.77% | Italy (3279) 7.28% | France (3215) 7.14% | Japan (3125) 6.94% | Netherlands (2776) 6.16% |
| MUL | USA (2897) 40.60% | England (1040) 14.57% | Germany (729) 10.22% | China (622) 8.72% | Australia (573) 8.03% | Saudi Arabia (514) 7.20% | France (501) 7.02% | Canada (467) 6.54% | South Korea (439) 6.15% | Japan (431) 6.04% |
| **PHY** | USA (23217) 31.75% | Germany (15557) 21.27% | South Korea (10849) 14.83% | France (10465) 14.31% | England (9757) 13.34% | China (9514) 13.01% | Italy (8722) 11.93% | Japan (8586) 11.74% | Russia (8580) 11.73% | Spain (6821) 9.33% |
| **SS** | USA (6928) 46.93% | England (3375) 19.98% | Australia (1777) 10.52% | Canada (1430) 8.47% | China (1398) 8.28% | Germany (1031) 6.10% | Netherlands (990) 5.86% | France (876) 5.19% | Switzerland (841) 4.98% | South Africa (724) 4.29% |

**Note:** Some research papers may be tagged in more than one subject area, due to overlaps in WC field. Similarly, an internationally collaborated paper may involve multiple countries.

Now, we try to identify which countries are the major collaborators in different subject areas. For this purpose, we analyze the collaborated publication records in each subject area and identify what are the major collaborating partners. **Table 3** presents the top 10 collaborating countries (arranged in descending order of collaborated papers) for all the 14 subject areas. While USA is found to have the highest number of collaborated papers across all subject areas, the proportion of collaborated papers vary from 47.93% in MED to 20.42% in MAR subject area. The other major collaborating countries in each subject areas are found to be somewhat different from each other. In AGR, other than USA, the other major collaborators are Australia, China, Germany and South Korea. In AH, the major collaborators (in addition to USA) are England, China, and Australia. In BIO subject area, the other major collaborators are England, Germany and South Korea; whereas in CHE, these are South Korea, Germany and Japan. South Korea, China, England and Germany are the other major collaborators in ENG subject area. In MED, the collaborating countries are England, Australia and Canada. In MAT subject area the other major collaborating countries are China, Saudi Arabia and Germany; whereas in PHY subject area, these are Germany, South Korea and France. Thus, the international collaboration in different subject areas show different patterns. Countries with more developed research in a subject area are likely to be among the major collaborating partners in that subject area. The existence of bilateral and multilateral cooperation programs in specific thematic areas may also be related to the international collaboration with a specific country. The different proximity dimension may also have a role to play in shaping subject area specific collaboration patterns with a set of specific countries. This is discussed further in the 'Discussion' section.

*RQ4: What proportion of internationally collaborated papers from India have lead/ majority author(s) from India?*

We now attempt to answer the next research question about the nature of international collaboration that Indian researchers are found to engage in. For this purpose, we compute two results- (a) how many internationally collaborated research papers have an Indian researcher as first/ lead author, and (b) what proportion of the internationally collaborated papers have 50% or more authors from India. The objective behind computing both these results is to find out if Indian researchers are playing the leading/ major role in the internationally collaborated papers they engage in or its otherwise.

**Table 4** presents the year-wise number and percentage of internationally collaborated papers that have an Indian researcher as first author, and also the proportion of papers that have at least 50% authors from India. It can be observed that the proportion of internationally collaborated papers with an Indian researcher as first author has increased marginally in the period, reaching to 52.15% in 2020 from 47.88% in 2001. From 2013 onwards, all the values are more than 50%. Thus, about half of the internationally collaborated papers from India during the period had an Indian researcher as first author. This indicates that more than half of the internationally collaborated research papers of India have a major role of research design and conceptualization played by the Indian researcher. Similarly, when we measure the quantum of Indian internationally collaborated papers with at least 50% Indian authors, we observe that more than 50% internationally collaborated papers have at least 50% author contribution from India. Although, it is also found that this proportion has decreased from 66.86% in 2001 to 51.12% in 2020. This is a significant drop in percentage value. There could be multiple reasons for the decrease, such as shift towards more team-oriented science, mobility of Indian scientists etc. However, at the same time, it is important to note that while proportion of research papers with at least 50% authors from India may have decreased, the proportion of papers with an Indian researcher as the lead author has increased marginally during this period.

The patterns of leading and at least 50% Indian authors have also been computed for research publications in different subject areas. **Table 5** presents the subject area-wise proportion of research papers with an Indian researcher as first author, and the proportion of research papers that have at least 50% authors from India. It is observed that, first author proportion lies between 36.15% (for SS subject area) to 60.24% (for CHE subject area). The subject areas with a good proportion of internationally collaborated research papers having an Indian lead author are: CHE (60.24%), MAR (56.84%), ENG (55.41%), INF (54.18%), BIO (51.03%) and PHY (50.00%). The subject areas with lesser proportion of internationally collaborated papers having Indian lead author are SS (36.15%), MED (40.47%), MUL (45.18%) and AGR (47.26%). In terms of internationally collaborated papers with at least 50% Indian authors, it is observed that subject areas MAT (64%), CHE (63.96%), ENG (63.66%) and MAR (63.46%) have higher proportion of internationally collaborated papers with at least 50% Indian authors. Many of these subject areas are also the one having higher proportion of research papers with Indian lead author. The subject areas of MED (44.65%), SS (47.54%) and MUL (47.69%) have less than 50% of research papers having at least 50% Indian authors. Thus, the patterns of lead as well as majority Indian authors in the internationally collaborated research papers is found to vary across different subject areas.

**Table 4: Authorship patterns in internationally collaborated papers**

| Year | No of internationally collaborated papers | Collaborated papers with Indian first author (%) | Collaborated papers with at least 50% Indian authors (%) |
|---|---|---|---|
| 2001 | 2354 | 1127 (47.88%) | 1574 (66.86%) |
| 2002 | 2707 | 1322 (48.84%) | 1780 (65.76%) |
| 2003 | 2986 | 1511 (50.60%) | 2018 (67.58%) |
| 2004 | 3409 | 1642 (48.17%) | 2184 (64.07%) |
| 2005 | 3838 | 1894 (49.35%) | 2426 (63.21%) |
| 2006 | 4623 | 2242 (48.50%) | 2975 (64.35%) |
| 2007 | 5188 | 2545 (49.06%) | 3351 (64.59%) |
| 2008 | 4810 | 2329 (48.42%) | 2961 (61.56%) |
| 2009 | 5784 | 2799 (48.39%) | 3554 (61.45%) |
| 2010 | 8089 | 3758 (46.46%) | 4913 (60.74%) |
| 2011 | 9071 | 4270 (47.07%) | 5483 (60.45%) |
| 2012 | 10018 | 4716 (47.08%) | 5785 (57.75%) |
| 2013 | 11181 | 5658 (50.60%) | 6274 (56.11%) |
| 2014 | 12807 | 6467 (50.50%) | 7018 (54.80%) |
| 2015 | 13771 | 6972 (50.63%) | 7386 (53.63%) |
| 2016 | 16412 | 8298 (50.56%) | 8832 (53.81%) |
| 2017 | 17746 | 9254 (52.15%) | 9508 (53.58%) |
| 2018 | 19871 | 10165 (51.15%) | 10478 (52.73%) |
| 2019 | 22018 | 11327 (51.44%) | 11601 (52.69%) |
| 2020 | 23815 | 12420 (52.15%) | 12175 (51.12%) |

**Table 5: Subject area-wise authorship patterns of internationally collaborated papers**

| Subject Area | No of papers in subject area[#] | Collaborated papers with Indian first author (%) | Collaborated papers with 50% Indian authors (%) |
|---|---|---|---|
| AGR | 11700 | 5530 (47.26%) | 6563 (56.09%) |
| AH | 1557 | 734 (47.14%) | 863 (55.43%) |
| BIO | 47080 | 24026 (51.03%) | 25738 (54.67%) |
| CHE | 46306 | 27896 (60.24%) | 29616 (63.96%) |
| ENG | 25656 | 14217 (55.41%) | 16333 (63.66%) |
| ENV | 17641 | 8554 (48.49%) | 9753 (55.29%) |
| GEO | 16394 | 8008 (48.85%) | 9047 (55.18%) |
| INF | 16031 | 8685 (54.18%) | 9439 (58.88%) |
| MAR | 42131 | 23948 (56.84%) | 26736 (63.46%) |
| MAT | 16371 | 7978 (48.73%) | 10478 (64.00%) |
| MED | 45058 | 18235 (40.47%) | 20117 (44.65%) |
| MUL | 7136 | 3224 (45.18%) | 3403 (47.69%) |
| PHY | 73135 | 36570 (50.00%) | 40182 (54.94%) |
| SS | 16893 | 6106 (36.15%) | 8031 (47.54%) |

[#] - Papers may be tagged in more than one subject area, due to overlaps in WC field.

### *RQ5: Do internationally collaborated papers of India get significantly higher impact and visibility as compared to indigenous papers?*

Many previous studies have highlighted advantage of international collaboration in terms of higher impact and visibility of research papers. Therefore, we tried to find out whether this holds true in case of Indian research papers involving international collaboration. For this purpose, we computed the citation impact as well as social media visibility of internationally collaborated and domestic research papers from India.

**Table 6** presents the cited % and citations per paper for internationally collaborated and domestic research papers. It can be observed that 87.68% of the total internationally collaborated papers get at least some citation as compared to 85.55% of the domestic papers. Thus, there is a slight edge in terms of citation potential seen for the internationally collaborated papers. Similarly, in terms of citations per paper, the value for internationally collaborated papers is 21.46 citations per paper as compared to 14.28 citations per paper for domestic papers. Thus, there is an edge in citations per paper as well for the internationally collaborated papers.

**Table 7** presents the social media visibility of internationally collaborated and domestic papers, in different social media platforms. It is observed that ICP instances have in general higher visibility as seen in terms of coverage by Altmetric.com aggregator, with 36.93% coverage for ICP and 19.45% coverage for Non-ICP papers. In Twitter platform, ICP instances get 23.28% coverage as against 12.23% for Non-ICP instances. The average tweets per paper for ICP and Non-ICP instances are 11.47 and 3.99, respectively. The Facebook and Blog platforms also show similar differentiation, with ICP instances getting higher coverage and visibility as compared to Non-ICP instances. Thus, the internationally collaborated papers from India are found to attract higher social media visibility as compared to domestic papers. Though, one must keep in mind the large difference in set of internationally collaborated and domestic papers while reading the summary/ average values. We can, however, understand that internationally collaborated research papers from India, in general, get an edge in citation impact and social media visibility as compared to domestic research papers.

**Table 6: Citation impact of ICP and domestic papers**

| Citation Impact | | |
|---|---|---|
| | ICP | Non-ICP |
| No of collaborated papers | 200,498 | 583,659 |
| Cited % | 87.68 | 85.55 |
| Citations per paper | 21.46 | 14.28 |

**Table 7: Social Media visibility of ICP and domestic papers**

| Social Media Visibility | | |
|---|---|---|
| | ICP | Non-ICP |
| No of papers | 200,498 | 583,659 |
| Altmetric Coverage % | 36.93 | 19.45 |
| Twitter Coverage% | 23.28 | 12.23 |
| Tweets per paper | 11.47 | 3.99 |
| FB Coverage% | 4.95 | 2.14 |
| Average FB mentions | 2.23 | 1.68 |
| Blog Coverage% | 3.05 | 0.76 |
| Average Blog mentions | 1.91 | 1.27 |

Now we discuss the subject area-wise citation impact of international collaborations of India. **Table 8** shows the subject area-wise overall impact of ICP and Non-ICP instances of India for the period 2001-2020. Overall impact (consolidated for the 20 years) is expressed as Cited % and Citations per paper (CPP). It can be seen from the table 8, that CHE (90.76%), AGR (89.90%) and BIO (89.54%) are three subject areas with relatively higher cited% in ICP, and subject areas such as CHE (90.73%), BIO (89.72%) and MAR (88.43%) have higher cited% in Non-ICP instances. Thus, few subject areas have higher cited % for all types of papers. Out of all 14 subject areas, AH has the lowest cited% in ICP and Non-ICP instances with 75.47% and 61.99% respectively. MUL (25.89), MED (25.78) and GEO (24.88) are the top subject areas having high CPP in ICP, whereas CHE (18.08), BIO (17.57) and GEO (17.20) are the subject areas having higher CPP in Non-ICP. Subject areas in which international collaborations made significant impact in comparison to domestic collaborations can be determined by CPP ratio, which is also shown in table 8. While the international collaborations are found to have significantly more impact than domestic collaborations in the subject area MUL, impact of international collaborations is found to be exactly double as that of domestic collaborations in the case of AH. For areas MED and PHY, it is also close to double.

**Table 8: Subject area wise citation impact**

| Subjects | ICP | | Non-ICP | | CPP ratio |
| --- | --- | --- | --- | --- | --- |
| | Cited % | Citations per paper | Cited % | Citations per paper | (CPP of Non-ICP/CPP of ICP) |
| AGR | 89.90 | 23.14 | 85.20 | 16.80 | 0.73 |
| AH | 75.47 | 9.66 | 61.99 | 4.83 | **0.5** |
| BIO | 89.54 | 22.69 | 89.72 | 17.57 | 0.77 |
| CHE | 90.76 | 20.86 | 90.73 | 18.08 | 0.87 |
| ENG | 86.29 | 17.91 | 84.30 | 14.54 | 0.81 |
| ENV | 88.91 | 22.57 | 87.09 | 16.94 | 0.75 |
| GEO | 89.77 | 24.88 | 87.34 | 17.20 | 0.69 |
| INF | 81.29 | 15.48 | 75.07 | 10.46 | 0.68 |
| MAR | 88.88 | 18.36 | 88.43 | 16.04 | 0.87 |
| MAT | 80.43 | 11.43 | 78.03 | 9.69 | 0.85 |
| MED | 88.37 | 25.78 | 86.29 | 13.99 | **0.54** |
| MUL | 83.16 | 25.89 | 73.24 | 8.75 | **0.34** |
| PHY | 88.98 | 21.67 | 86.05 | 12.20 | **0.56** |
| SS | 85.53 | 16.77 | 80.39 | 10.78 | 0.64 |

**Discussion**

The paper analysed the research publication data of India during 2001 to 2020 to measure and characterize the international collaboration patterns. It is observed that the proportion of India's internationally collaborated papers has grown significantly during the last 20 years, rising from 20.73% in 2001 to 32.35% in 2020. The CAGR value of internationally collaborated papers is 12.27%, which is higher than the growth rate of total research publications from India (9.80%). Given that international collaboration in research is known to have several advantages, it is advantageous for India to be able to engage in more internationally collaborated research. When compared to international collaboration levels with other developed countries, we find that India stands between the high ICP level of England and lower ICP level of China, among the major research output producing countries. While, England has a high proportion of 57.7% of its research papers as ICP instances, China has 20.3% its research papers as ICP instances, as in 2016 (Singh et al., 2021). Among other major knowledge producing countries, USA has 36.8% of its research papers internationally collaborated, whereas Germany has 50.5% of its research papers that are internationally collaborated. Thus, in terms of proportion of international collaboration in research, India is engaging in more internationally collaborated research as compared to China, is closer to USA, but much less than Germany and England. The growth in research output from India may also have some connection with rise in international collaboration, as has been observed in (Glanzel, 2001; Abramo, DÁngelo, & Solazzi, 2011; Abramo, DÁngelo & Murgia, 2017). Countries which have developed international collaboration in research have gained in terms of productivity as well as impact, and India may be no different.

*International Cooperation Programs and their role*

The major international collaborating partners for India, during the 2001 to 2020 period, are USA, Germany, England, South Korea and China. Other important emerging collaborators are Saudi Arabia, Australia, South Africa, Malaysia, and France. It may be noted that India is participant in many bilateral and multilateral international cooperation programs[3] involving several countries, which could also have played some role in development of research collaboration with some of these countries. For example, India has extensive collaboration program in different areas with USA, UK, Germany, France etc. There are in fact dedicated organizations/ centers working to promote cooperation and collaboration programs with these countries. These centers include Indo-US Science and Technology Forum[4], UK-India Education and Research Initiative[5], Indo-German Science and Technology Centre[6], Indo-French Centre for the Promotion of Advanced Research[7] etc. Recently under its look east policy, more cooperation programs have been instituted with South East Asian countries like South Korea, Philippines, Malaysia etc. However, unfortunately the association of India's neighboring South Asian countries in the SAARC intergovernmental organization involving many kinds of cooperation have not been very fruitful in promoting research collaboration in these countries, as India's international collaboration with its South Asian neighbors is found to be extremely small. Nevertheless, bilateral and international cooperation programs have a definite role to play in promoting international research collaboration between countries. This is also evident from the fact that out of total internationally collaborated research papers from India, about 67.21% research papers are examples of bilateral collaboration. However, at the same time the proportion of multilateral collaboration is seen rising during this period. The empirical observations in the paper can thus be used to assess the effectiveness of different cooperation programs of India to promote research collaboration. The lessons learned from the data and the past experiences can be used to modify the existing programs and to institute new international cooperation programs in science and technology.

---

[3] https://dst.gov.in/international-st-cooperation
[4] https://iusstf.org/
[5] http://ukieri.org/
[6] https://www.igstc.org/
[7] http://www.cefipra.org/

*Role of Proximity Dimensions*

One may also like to see the international collaboration patterns in the light of various proximity dimensions. Many previous studies have explored role of different proximities/ distance dimensions as facilitators/ inhibitors of international collaboration (Boschma,2005; Balland, Boschma & Frenken, 2015; Vieira, Cerdeira, & Teixeira, 2022). The conceptualization of proximities and their role in scientific collaboration has been discussed in several previous studies (Kraut et al., 2002; Caniels, Kronenberg, & Werker, 2014; Fernandez, Ferrandiz, & Leon, 2016; Heringa, Hessels & van der Zouwen, 2016). The following distance/ proximity dimensions are mainly analysed in the previous studies: geographical (Katz, 1994; Kuhn, 1996; Nagpaul, 2003; Ponds, Van Oort, & Frenken, 2007; Hoekman, Frenken & Tijssen, 2010; Laursen, Reichstein, & Salter, 2011; Huang, Shen & Contractor, 2013; Wagner & Jonkers, 2017, Berge, 2017), socioeconomic (Sokolov-Mladenovic, Cvetanovic, & Mladenovic, 2016; Fernandez, Ferrandiz, & Leon, 2016; Jiang et al., 2018), political (Boschma, 2005; Cattaneo & Corbellini, 2011; Dawes, Gharawi, & Burke, 2012; Whetsell et al., 2020), Cultural (Knoben & Oerlemans, 2006; Narteh, 2008; Hwang, 2013; Plotnikova & Rake, 2014; Suhay & Druckman, 2015; Taylor & Osland, 2015; Gui, Liu, & Du, 2019), intellectual (Cohen & Levinthal, 1990; Powell, Koput, & Smith-Doerr, 1996; Cummings & Teng, 2003; Hoekman, Frenken & Tijssen, 2010), and excellence (Jones, Wuchty, & Uzzi, 2008; Abramo, DÁngelo & Solazzi , 2011; Jeong, Choi, & Kim, 2014; Abramo, DÁngelo, Murgia, 2017). In general, it has been observed that different kinds of proximities may facilitate international collaboration, however, the degree of impact of different proximity dimensions vary in different situations. Recently, Hou, Pan & Zhu (2021) reframed the major factors and explored the impact of scientific, economic, geo-political and cultural factors on international research collaboration. They found that countries with large and equivalent scientific sizes, as well as economic sizes, are more likely to collaborate closely with each other. Co memberships in intergovernmental organizations and cultural links (such as shared language and religion) also facilitate close collaboration between countries.

In case of Indian international collaboration, out of geographical, socio-economic, political, cultural, intellectual and excellence proximities; the intellectual and excellence dimensions appear to be more effective. Most of the major collaborating countries for India are located quite far geographically and have different cultural settings. In case of England, political and cultural dimensions may be effective since India and England share similar political systems and common language used by academicians and researchers. Intellectual and Excellence dimensions may explain India's higher international collaboration with USA, Germany, South Korea, Saudi Arabia, Australia etc. India's international collaboration with its South Asian neighbors is extremely small to be noticed, therefore the geographical, socio-economic and cultural proximity dimensions do not appear to be playing a major role in this respect. These observations are, however, only indicative in nature and a more systematic procedure and experimental design will be required to confirm the impact of various proximity dimensions on India's international collaboration patterns.

*Knowledge Flow in Indian International Research Collaboration*

The authorship structure of internationally collaborated papers from India indicate that Indian authors play major role in such collaborated papers. The number of papers with Indian lead (first) author has increased during 2001 to 2020. Indian researchers are the lead authors in more than 50% of the internationally collaborated research papers. Similarly, more than 50% internationally collaborated research papers have at least 50% authors from India. Both these patterns together demonstrate that Indian researchers have played major role in the internationally collaborated research papers. This in turn may be an indication of the knowledge flows and important role played by Indian researchers in such collaborated research. The capability of Indian researchers in fostering international collaborations seem to have improved over the time. The more noticeable observation here is that this improvement is not at the cost of importance of their contribution, as Indian authors are found to be in leading position in the rising international collaboration. The role and contribution of Indian researchers in such collaborated output, however, is also found to varying in different subject areas, with some subject areas having more than 50% research papers with Indian lead author. The subject areas of CHE (60.24%), MAR (56.84%), ENG (55.41%), and INF (54.18%) have good proportion of the internationally collaborated papers with Indian lead author. This indicates that the international collaboration in the areas of Chemistry, Materials, Engineering and Information Technology have a more impactful role of Indian researchers in the collaboration. The subject areas like SS (36.15%), MED (40.47%), MUL (45.18%) and AGR (47.26%) subject areas have lesser proportion of such papers, indicating inward knowledge flow in these areas. Interestingly, SS, MED and MUL are also the subject areas with less proportion of internationally collaborated papers having at least 50% authors from India. It appears that the subject areas where India may have a strong research publication base may have a more impactful participation of Indian researchers in international collaboration. This information may be used to institute subject/ theme specific collaboration programs on a complementing basis.

*Citation and altmetric impact of internationally collaborated research*

The international collaboration advantage in research impact and visibility is also seen in Indian research output as measured in terms of citations and social media visibility. Several previous studies have demonstrated such advantages (Bordons et al., 1996; Glanzel & Schubert, 2001; Persson, 2010; Bote et al, 2013; Nguyen et al., 2017). In case of India, a marginal advantage in cited% is seen for internationally collaborated papers as compared to domestic papers. The citations per paper value for internationally collaborated papers is higher than domestic papers. There are subject area-wise variations observed too. The advantage is more profound in case of social media visibility. Both, the social media coverage and average mentions per paper is significantly higher for internationally collaborated papers as compared to domestic papers. Several reasons could be responsible for this. An internationally collaborated paper gets more attention across research groups in different countries and hence is likely to be more impactful in terms of citations and social media visibility. Further, internationally collaborated papers may be addressing research problems of wider interest across national boundaries and hence they may be able to attract more attention.

*Some Policy Implications*

As international collaborations have turned out well for India in terms of both social media visibility and citation impact in comparison to domestic collaborations (on exploration based on RQ5), we highly recommend Indian institutions to strengthen the efforts to engage in more international collaborations and maintain existing collaborations. Based on the results of investigation carried out for RQ3, such effort can be directed by giving more emphasis to subject areas such as Agriculture Science (AGR) and Engineering (ENG) where international collaboration is relatively weak. Though international collaborations are highly rewarding, considering the cost of establishing and maintaining such collaborations, barring some top-level institutions, most of the Indian institutions might be finding it hard to break the ice for that matter. So, it is high time Indian institutions made deliberate efforts to make domestic collaborations as successful in terms of intensity, citation impact and visibility on par with the international collaborations. As determination of right partner for collaboration is very much vital for successful collaborations, one of the major reasons for finding domestic collaborations to be less rewarding than international collaborations can be the choice of partners of collaboration. At the level of institutions, to engineer collaborations with compatible institutions that can complement their expertise, institutions should have an effective mechanism to identify their research strengths, i.e., the areas in which they are really strong (core expertise or core competency areas) and the areas that have the potential to develop as a core competency area later (potential core competency areas). For this, they can adopt any sound assessment methods that help to reveal the research portfolio. Internal audit or assessment can be useful to an extent for this. But in order to identify suitable partners for an institution, it is important to know the strengths of other institutions too and such information is difficult to obtain. For policymakers at national level too, this exercise can be vital. This approach can be used for performance-based funding/resource allocation decisions in 'thrust-areas' or areas of national importance.

Apart from this, the same approach can be useful for capacity building for systematic institutionalization for nurturing institutional clusters of excellences in these thrust-areas to make the country globally competitive in these areas. Lathabai, Nandy and Singh (2021) developed a framework to determine research portfolio (that includes core competency areas and potential core competency areas) of academic institutions of a country and Lathabai, Nandy and Singh (2022) developed a recommendation system framework that can toss collaboration recommendations based on the core competency and potential core competency areas for institutions with respect to a field. Application of these frameworks can be extended beyond the limit of fields to explore suitable partner institutions for collaboration in multiple thematic areas. For that matter, one contribution of this research can be significant. As mentioned earlier, agriculture science (AGR) and engineering (ENG) are found to be the subjects with more domestic collaborations and from CPP ratio of these subject areas (0.73 and 0.81 respectively) domestic collaborations are found to be fairly rewarding in comparison to international. Thus, as there are signs of existence of impactful domestic collaborations, further explorations for establishment of more impactful domestic collaborations should be conducted. Considering this and the strategic importance of these fields for the nation, along with attempting to improve the level of international collaborations, systematic approach for bringing out impactful domestic collaborations with respect to important fields within these subjects by application of the above discussed recommendation system framework will be really valuable at institutional level as well as to the country as a whole.

As areas MUL, AH, MED and PHY are the areas where international collaborations are found to be more rewarding, efforts for maintaining such collaborations can be encouraged. However, does this point towards the weakness of domestic collaborations? Though an ideal value of the CPP ratio to indicate the alarming lack of self-sufficiency in a subject-area is neither given nor any rule of thumb is defined to do so in the available literature, values of CPP close to 0.5 and less can be taken as a warning signal. Special care should be taken in these subject areas to make domestic collaborations more impactful. Thus, this study can serve as a basis for prioritization of subjects (from which important fields can be picked up). Therefore, another recommendation to national level policymakers is that while considering key fields for fostering collaborations and to build institutional clusters of excellence, give special emphasis to the fields within MUL, AH, MED, PHY, AGR and ENG. Since the major collaborating countries for different subject areas are found to be a little different, subject-specific strategies for international collaboration can be worked out. For example, England is the major collaborator in subject areas- SS, MUL, MED, BIO and GEO. China is major collaborator in ENV and INF subject areas. Similarly, Germany is major collaborator in PHY subject area, whereas South Korea is in CHE and Australia is in AGR subject area. Thus, the subject specific differentiations in international collaborations can be observed and utilized for instituting appropriate collaboration programs.

**Conclusion**

The article measures and characterizes the international collaboration patterns in Indian scientific research during the last two decades (2001-2020). It is found that Indian international collaboration in Indian scientific research has rose from 20.73% internationally collaborated papers in 2001 to 32.35% internationally collaborated papers in 2020. Among internationally collaborated papers, a large proportion (67.21%) constitutes case of bilateral collaboration. USA, Germany, England, South Korea and China are the major collaborating partners, though there are variations in the magnitude of collaborated research output across different subject areas. Multidisciplinary, Arts & Humanities, Medical Science and Physics subject areas are the ones to gain the most through international collaboration. Out of the internationally collaborated papers from India, more than 50% have an Indian researcher as lead author, and more than half of the papers have at least 50% authors from India. The internationally collaborated research output of India is also found to have an advantage in terms of citations and social media visibility as compared to domestic papers. Thus, the article not only answers the research questions proposed but also presents a comprehensive and updated analysis and characterization of international collaboration patterns in Indian scientific research, which can be useful for various purposes. The empirical observations in the paper can be used to assess the effectiveness of different cooperation programs of India to promote research collaboration. The lessons learned from the data and the past experiences can be used to modify the existing programs and to institute new international cooperation programs in science and technology. Some recommendations are made to policymakers at institutional and national level along with directions for further research that can be useful for the improvement of the overall academic performance of the country.

## Declarations

**Funding and/or Conflicts of interests/Competing interests:** The authors declare that no funding was received for this work. Further, the manuscript complies with ethical standards of the journal and there is no conflict of interests whatsoever.